# Cryogenic Characteristics of Graphene Composites – Evolution from Thermal Conductors to Thermal Insulators

Zahra Ebrahim Nataj,[1,*] Youming Xu,[2,*] Jonas Brown,[1] Jivtesh Garg,[3] Xi Chen,[2] Fariborz Kargar,[1,†] and Alexander A. Balandin[1]

[1]Phonon Optimized Engineered Materials Center, Department of Electrical and Computer Engineering, University of California, Riverside, California 92521 USA

[2]Department of Electrical and Computer Engineering, University of California, Riverside, California 92521 USA

[3]Department of Aerospace and Mechanical Engineering, University of Oklahoma, Norman, Oklahoma 73019 USA

[*] These authors contributed equally to the manuscript.
[†] Corresponding authors: fkargar@ece.ucr.edu ; balandin@ece.ucr.edu

# ABSTRACT

The development of cryogenic semiconductor electronics and superconducting quantum computing requires composite materials that can provide both thermal conduction and thermal insulation. We demonstrated that at cryogenic temperatures, the thermal conductivity of graphene composites can be both higher and lower than that of the reference pristine epoxy, depending on the graphene filler loading and temperature. There exists a well-defined cross-over temperature – above it, the thermal conductivity of composites increases with the addition of graphene; below it, the thermal conductivity decreases with the addition of graphene. The counter-intuitive trend was explained by the specificity of heat conduction at low temperatures: graphene fillers can serve as, both, the scattering centers for phonons in the matrix material and as the conduits of heat. We offer a physical model that explains the experimental trends by the increasing effect of the thermal boundary resistance at cryogenic temperatures and the anomalous thermal percolation threshold, which becomes temperature dependent. The obtained results suggest the possibility of using graphene composites for, both, removing the heat and thermally insulating components at cryogenic temperatures – a capability important for quantum computing and cryogenically cooled conventional electronics.

**KEYWORDS**: cryogenic electronics; thermal management; thermal conductivity; graphene composites; thermal percolation; quantum technologies

There is a rapidly emerging need for thermal management at cryogenic temperatures. It is driven by several trends and developments. There is a strong motivation to run conventional semiconductor electronics at low temperatures to implement "cold computing," which allows one to increase computational and energy efficiency while reducing power consumption.[1,2] The main progress with quantum computing technologies is associated with superconducting qubits, which require cryogenic temperatures.[3] Space exploration needs electronics operating in harsh low-temperature environments. Cryogenic thermal management usually relies on both thermal conductors and thermal insulators.[4–6] The former is the polymer-based thermal interface materials (TIMs) with fillers that conduct heat well, facilitating heat removal, and the latter are polymer materials, which have low thermal conductivity and can act as thermal insulators between electronic components operating at different temperatures. Examples of low-temperature thermal management applications include protective coatings of superconductive power cables[7], adhesives in cryogenic low-noise amplifiers for radio-astronomy and space communication systems,[8] optical mounts of cryogenic refractive optics and cryo-sorption pumps.[9,10]

Polymers are poor heat conductors with thermal conductivity in the range from ~0.2 to $0.5 \ \mathrm{Wm^{-1}K^{-1}}$ at room temperature (RT).[11,12] Polymers are used in TIMs as the base, *i.e.* matrix material, that fills the air gaps between two adjacent solid interfaces and provides adhesive functionality when used in curing composites. The common strategy to increase the thermal conductivity of polymers is to add micrometer- and nanometer-scale fillers with a higher intrinsic thermal conductivity that can couple well with the base polymer. A mixture of the single-layer and few-layer graphene flakes, termed "graphene" in the thermal context, has proven to be an efficient filler material for a variety of TIMs, including non-curing mineral oil-based thermal pastes,[13] and curing epoxies.[12,14,15] Graphene for thermal management applications can be mass-produced *via* liquid-phase exfoliation, graphene oxide reduction, or other techniques.[16,17] Graphene TIMs with thermal conductivity above $\sim 12 \ \mathrm{Wm^{-1}K^{-1}}$ near RT, which exceeds the metric of conventional commercial TIMs, have been reported by several research groups.[12,14,15] The excellent performance of graphene TIMs near RT originates in the extraordinarily high intrinsic thermal conductivity of graphene and few-layer graphene,[18,19] strong coupling to the matrix, good dispersion, and appropriate viscosity range of the resulting composites.[20] The enhancement of thermal conductivity is achieved both below and above the thermal percolation threshold – a loading fraction at which the graphene fillers start to form continuous thermally conductive paths.[21,22] The thermal percolation threshold can be identified when the dependence of the thermal conductivity on filler loading becomes super-linear.[21,23] One should note that the thermal properties of graphene composites have only been studied at RT and above – the temperature range of interest for conventional electronics. We are not aware of any report on the cryogenic thermal characteristics of graphene composites. In general, the data for the thermal properties of any polymer composites at cryogenic temperatures are scares. The understanding of heat propagation in amorphous polymers at low temperatures is far from complete, even if one does not consider the solid inclusions, *i.e.* fillers.[24,25]

Here, we investigated the thermal properties of epoxy–graphene composites at temperatures from 2 K to RT. Epoxy is a practically important material and it is often used as a reference material to compare the effect of different types of fillers on its thermal conductivity. We found that at cryogenic temperatures, the thermal conductivity of graphene composites can be both higher and lower than that of the reference pristine epoxy, depending on the graphene filler loading and temperature. This is drastically different from what is observed near RT. Moreover, there exists a well-defined cross-over temperature that above it, the thermal conductivity increases with the addition of graphene, whereas, below it, the thermal conductivity decreases with the addition of graphene. Graphene composites are unique in a way that they can provide both the strongest enhancement in thermal conductivity and the strongest suppression. We offer a physical model explanation of the counter-intuitive trends and provide numerical simulation data which agree with the measurements. The obtained results suggest the possibility of using composites with the same constituent materials for, both, removing the heat and thermally insulating electronic components at cryogenic temperatures. The latter constitutes a *conceptual* change for thermal management, which typically rely on different materials for heat conduction and isolation.

## EXPERIMENTAL SECTION

**Materials:** The polymer matrix is a thermoset epoxy set consisting of a base resin (Bisphenol-A, Allied High-Tech Products, Inc.) and a hardening agent (Triethylenetetramine; Allied High-Tech Products, Inc.). We used few-layer graphene with the specified average lateral dimension of $25\ \mu m$ and an average surface area of $50$ to $80\ m^2 g^{-1}$ (xGnP, Grade H, XGSciences, the US) as the fillers for the preparation of the composites. The graphene fillers were subjected to further exfoliation during the high-speed mixing process with the base polymer matrix. The resulting fillers constitute a mixture of FLG and single-layer graphene. The lateral dimensions of graphene fillers are an important parameter for tuning the thermal conductivity of the composites.[26] To maximize the thermal conductivity of the composite, one normally wants to keep the lateral dimensions above the gray phonon mean free path (MFP) in graphene, which is about ~1 μm near RT.[27] In contrast, to improve the insulating properties of composites, fillers with smaller lateral dimensions are preferred.[28]

**Composite Preparation and Characterization:** Several composite samples were prepared by mixing precalculated quantities of the resin, the hardener, and the FLG fillers to hit a targeted filler loading level. First, a certain amount of FLG fillers was distributed into the epoxy resin using a high-speed shear mixer (Flacktek, Inc.). The hardener was then added to the mixture in the mass ratio of 12:100 with respect to the base epoxy resin's weight. The blend was vacuumed several times to eliminate any air bubbles that might have been trapped during the composite preparation process. The mixture was poured into special molds to cure and solidify. The results of the mass density measurements confirms that the porosity of the samples is negligible (Figure S1).The full recipe of the preparation is detailed in the Supplemental Information. An optical image of representative samples with various graphene loadings is presented in Figure 1 (a). Figure 1 (b) is an SEM image of a composite sample with the filler concentration of $f = 21.7\ \mathrm{vol}\%$. As the volume fraction of fillers increases beyond a certain loading, referred to as the percolation threshold, the fillers start to overlap (see SEM image in Figure 1 (b)). At and beyond the percolation regime, fillers create a network of electrically and thermally conductive pathways within the base polymer matrix.[21,22,29] The percolation results in significant enhancements in both the electrical and thermal characteristics of the composites.[21,22,29] The composites were further characterized using Raman spectroscopy (Renishaw InVia) with the laser excitation wavelength of $\lambda = 633$ nm in the conventional backscattering configuration. In all experiments, the laser power was kept at ~3 mW. Raman spectra of several samples at random spots are presented in Figure 1(c). In all composites, the characteristic G-peak and 2D-peak of graphene were observed at ~1580 cm$^{-1}$ and ~2700 cm$^{-1}$ [Ref. [30]] confirming an even distribution of graphene throughout the samples. The intensity of the 2D peak is much lower than the G-peak indicating the presence of a mixture of single-layer and few-layer graphene in the composite samples.

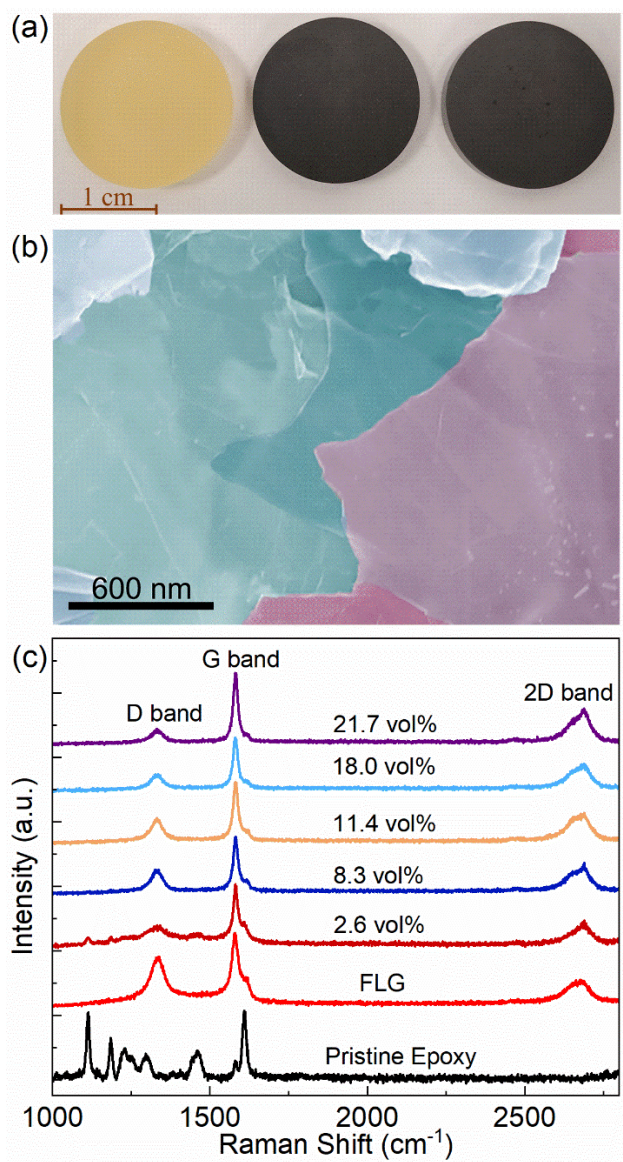

[**Figure 1: Sample preparation and characterization.** (a) From left to right, optical images of pristine epoxy and composites with 2.6 vol% and 5.4 vol% loading of few-layer graphene. (b) SEM image of an epoxy sample with 21.7 vol% graphene loading. Note the regions where the fillers overlap. (c) Raman spectra of pristine epoxy, few-layer graphene, and composites with various graphene concentrations.]

## RESULTS AND DISCUSSIONS

**Specific heat of graphene composites at cryogenic temperatures:** We start with the investigation of the specific heat and illustrate its typical temperature trends in disordered amorphous materials. Figure 2 (a) and (b) show the results of the specific heat, $c_p$, measurements of pristine epoxy and selected composites as a function of temperature for different filler concentrations. The data are presented in two different temperature ranges to make the trend more explicit (also, see Figure S2). The shaded area around the experimental data points indicates the standard error in the measurements conducted with Physical Property Measurement System (PPMS, DynaCool, Quantum Design). The results for the neat epoxy agree well with the literature.[31] The variation of the specific heat with temperature resembles that in other amorphous polymers.[31–33] In all samples, $c_p$ increases more rapidly in the low-temperature region; the rate of increase slows down as the temperature rises. To better visualize the temperature characteristics of $c_p$, we also plotted the data in a log-log scale, indicating the polynomial functional dependencies (Figure 2 (c)). One can distinguish three regions: quasi-cubic, $c_p \sim T^{3+\delta}$, in the temperature range of 2 K $\leq T \leq$ 6 K; parabolic, $c_p \sim T^2$, in the interval of 6 K $\leq T \leq$ 35 K; and linear, $c_p \sim T$, in the range of $T \geq$ 35 K. Among these regions, the $c_p$ behavior in the low-temperature limits is of particular interest since it deviates from the classical Debye model for crystalline materials. We plotted the "Debye-reduced" heat capacity, $c_p T^{-3}$, as a function of temperature in a log-log scale in Figure 2 (d). In the insulating crystalline materials, $c_p \sim T^3$, and therefore one expects to see a flat curve in the low-temperature limits of the $c_p T^{-3}$ *vs.* $T$ plot. For the pristine epoxy and composites with graphene, there is no $T^3$ dependency of $c_p$. The latter stems from the amorphous, disordered natures of these materials.

One can notice from Figure 2 (d) that as the graphene loading increases, the curves flatten out, getting closer to the $c_p \sim T^3$ characteristic of crystalline materials. It is well-established that the specific heat of amorphous materials exhibits a "universal" characteristic in the low-temperature limits.[34,35] Below $T \sim$ 1 K, the heat capacity considerably exceeds the Debye model predictions, dominated by a quasi-linear temperature dependence.[34,35] This behavior is explained by the tunneling model of two-level systems for amorphous materials at low temperatures.[36] This anomalous quasi-linear trend is followed by a hump in the $c_p T^{-3}$ data, referred to as the "boson peak" in the 3 K $\leq T \leq$ 10 K region.[36–38] The "boson peak" shown in Figure 2 (d) is attributed to the low-frequency vibrational modes present in amorphous materials.[38] The temperature at which the boson peak occurs decreases slightly with increasing the filler loading. The "boson peak" eventually fades away with increasing the filler content as the material acquires more "crystalline" characteristics owing to the graphene content. The overall value of $c_p$ decreases with increasing graphene loading since $c_p$ of FLG is much lower than that of the epoxy matrix. We established the general trends of the specific heat of graphene composites that follow those of amorphous materials but become more crystalline-like with increasing graphene content.

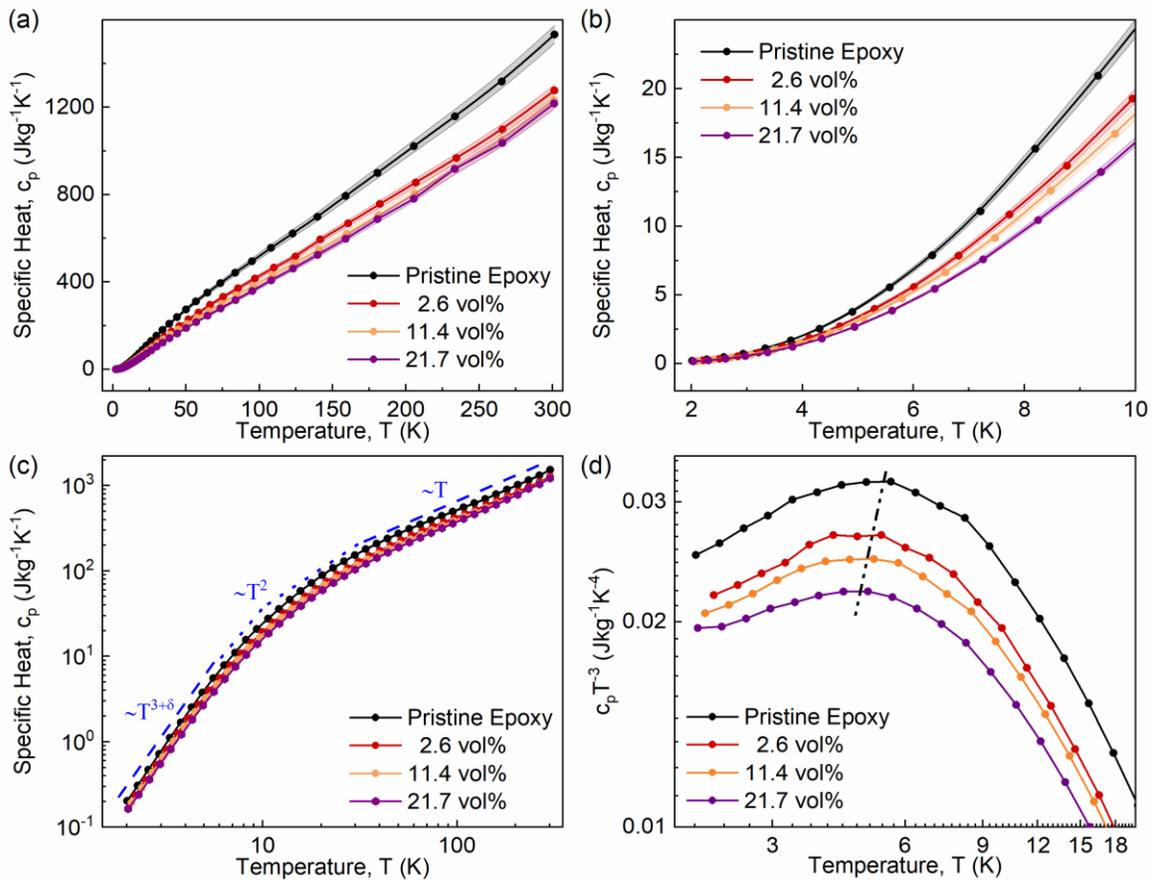

[**Figure 2: Specific heat of graphene composites.** (a) Temperature-dependent specific heat of composites in the temperature range of $2\ K \le T \le 300\ K$. (b) The same data as in panel (a) shown in the low-temperature limits. (c) Specific heat of the composites plotted in the log-log scale, revealing the quasi-cubic, parabolic, and linear temperature dependence in the low, intermediate, and high-temperature ranges. (d) The "Debye-reduced" specific heat of graphene composites as a function of temperature in a the log-log scale. The dashed line is a guide to the eye, showing the behavior of the so-called "boson peak" as a function of filler loading and temperature.]

**Thermal conductivity of graphene composites at cryogenic temperatures:** We now turn to the main topic of this study – the thermal conductivity of graphene composites at cryogenic temperatures. Figure 3 (a) shows the thermal conductivity, $k$, measured using PPMS in the temperature range from 2 K to RT in a log-log scale. The details of the measurements are described in the Supplemental Materials. The thermal conductivity of the pristine epoxy initially increases sup-linearly in the temperature range of $2 \text{ K} \leq T \leq 6$ K, followed by a "plateau region" in the interval of $6 \text{ K} \leq T \leq 17$ K, where $k$ remains nearly constant. After the plateau, $k$ increases linearly again until $T \sim 80$ K at which point a second plateau extending up to $T \sim 175$ K occurs. The existence of the first plateau in $k$ vs. $T$ dependence is universal for amorphous materials, and it occurs almost in the same temperature range. It starts at the temperature where the "boson peak" appears in the specific heat (see Figure 2 (d)).[39] The first plateau region is explained by the two-level systems and the tunneling model.[36,39] According to this model, the dependence is due to the cross-over at which phonons with shorter MFP become dominant heat carriers instead of phonons with longer MFP.[36,39] The thermal conductivity is given by $k = (1/3) \, Cv\Lambda$ in which $C$, $v$, and $\Lambda$ are the specific heat, average group velocity and the mean free path of phonons, respectively. In the plateau region, while $C$ increases with the temperature, $\Lambda$ decreases almost at the same rate. In other words, the plateau forms at the temperature range where the product of $\Lambda$ and $C$ becomes independent of temperature. The second plateau region was explained by similar considerations but its origin is still the subject of debate.[40]

Before proceeding further with the analysis of the experimental thermal data we should clarify the use of the "phonon" concept and terminology. Naturally, the introduction of phonons – quanta of crystal lattice vibrations – requires translational symmetry encountered in crystalline solids.[41] Amorphous materials lack translation symmetry. Complex models involving different descriptions of atomic vibrations referred to as propagons, diffusons, and locons, corresponding to propagating, diffusion, and localized modes, have been introduced to describe thermal transport in amorphous materials.[42,43] However, in the context of the present study of heat conduction in epoxy-based composites, we still can use the concept of acoustic phonons as the elastic vibrations in the continuous medium of the base material. The acoustic phonons are the dominant heat carriers in dielectric materials, including amorphous epoxy. We can talk about the acoustic phonons, *i.e.*

vibrations, propagating through the matrix that carry heat and interact with graphene fillers. For analysis of heat conduction in graphene composites, it is also useful to keep in mind the concept of the two-phase semi-crystalline medium where crystalline fillers interface with the amorphous polymer host.[34]

Let us now consider the temperature-dependent characteristics of the thermal conductivity of the graphene–epoxy composites in more detail. In Figure 3 (b), we replotted the thermal conductivity data for the same graphene loadings on a linear scale. There is no plateau region in graphene composites similar to the one observed for pristine epoxy. For composites with the filler loading $f \leq 8.3$ vol%, there is a well-defined cross-over temperature, $T_c \sim 11$ K, at which $\kappa$ is the same for the pristine epoxy and graphene composites with variable graphene loading. The existence of a cross-over temperature and the absence of a plateau have been reported previously in the thermal conductivity data of semi-crystalline materials[44] and polymer-based composites with filler inclusions, which can be considered semi-crystalline media.[34] For example, while the plateau in $k$ emerges in the amorphous polyethylene terephthalate (PET), it vanishes in semicrystalline PET containing crystalline zones with volume fraction as low as $f = 9$ vol%.[44] The $T_c \sim 15$ K is reported for PET with crystalline regions of $0 \leq f \leq 51$ vol% dispersed inside amorphous PET.[44] Several studies have reported a cross-over temperature for different composite systems in a similar temperature range. In all these studies, the plateau region vanished as soon as crystalline fillers were added to the amorphous polymer.[28,45–48]

The peculiar features in the temperature-dependent thermal transport characteristics can be explained by the thermal boundary resistance (TBR), $r_b$, at the interface of the amorphous polymer and crystalline FLG fillers.[28,44] According to the acoustic mismatch theory, at sufficiently low temperatures, $r_b \sim T^{-3}$, and thereby $r_b$ shows strong effects in the low-temperature limits.[49,50] The effect becomes small at high temperatures. Note that, below $T_c$, $\kappa$ of the pristine epoxy is larger than that of the epoxy composites with graphene fillers. This means that the inclusion of FLG fillers into the amorphous epoxy reduces its $k$ when the temperature is below $T_c$ but improves it when the temperature is above the cross-over temperature. This is a direct consequence of the strong temperature dependence of TBR at the polymer—filler interfaces. This will be further discussed in the theory section.

Figure 3 (c) shows the thermal conductivity of graphene composites as a function of the filler loading, $f$, at constant temperatures of 2 K, $T_c = 11$ K, and RT. The systematic errors for the data points at low temperatures were smaller than the size of the symbols and not shown for clarity. At $T = 2$ K, the thermal conductivity of the epoxy falls approximately two times with the addition of only 2.6 vol% of graphene fillers. The sharp decrease in thermal conductivity suggests that FLG fillers contribute significantly to the phonon scattering processes. At this temperature, as the concentration of the filler increases, the thermal conductivity declines linearly. At RT, however, the trend is the opposite. The addition of graphene fillers improves the thermal conductivity monotonically in line with many prior reports.[12,21,23] The variation of $k$ as a function of $T$ is more intriguing in the vicinity of the cross-over temperature, $T_c \sim 11$ K. With increasing the graphene filler loading, thermal conductivity remains the same up to $f = 8.3$ vol%; then, linearly increases in the interval of 8.3 vol% $\leq f \leq 14.6$ vol%; and after that, remains unchanged with adding more fillers. Typically, a nonlinear behavior such as the one observed at $T_c \sim 11$ K might indicate that the composite system at $f = 14.6$ vol% enters the thermal percolation regime. The absence of such nonlinear behavior at $T = 2$ K and RT suggests an intriguing possibility of the temperature-dependent percolation threshold.

Typically, one thinks about the percolation threshold in terms of the filler loading, $f_H$, as the point where the fillers start to mechanically touch each other, forming a continuous conductive network. Thermal percolation is less abrupt than electrical percolation because the heat can be conducted by the matrix materials, contrary to the electrical current that cannot be conducted by the dielectric matrix.[21,51,52] The ratio of the thermal conductivity of the fillers to the matrix, $k_f/k_m$, is several orders of magnitude smaller than the ratio of the electrical conductivity of the fillers to the matrix, $\sigma_f/\sigma_m$ .[21] The wavelength of the thermal phonons, which make the dominant contribution to heat conduction, is $\lambda_T \sim hV_s/k_bT$, where $h$ is the Plank's constant, $k_b$ is the Boltzmann's constant and $V_s$ is the sound velocity, or more accurately, the phonon group velocity.[34] For a typical semiconductor material $\lambda_T$ is on the order of 1 nm – 2 nm at RT.[53] This means that for the thermal percolation to occur at RT, the fillers should be in physical contact or close to each other, *i.e.,* on the order of $\sim$ 1 nm. At low temperatures, $\lambda_T$ increases by more than an order of magnitude. Thus, phonons with long wavelengths might provide thermal "cross-talk" to the fillers over some distance. The average distance between the fillers at $\sim$10 vol % is still larger than $\lambda_T$. However, one should remember that more refined theories of thermal conductivity attribute a more significant contribution to heat conduction to the phonons with wavelength above $\lambda_T$.[54] If we accept this picture of the process, then the dependence of the thermal conductivity at 2 K and 300 K are those below and above thermal percolation, correspondingly (see Figure 3 (c)). It is illustrative to analyze them further with the specific heat dependence on the loading fraction shown in Figure 3 (d). The specific heat decreases with increasing $f$ for both temperatures since $c_p$ of FLG is lower than that of the matrix. The thermal conductivity decreases at 2 K with $f$ either following the $c_p$ trend or because graphene fillers are acting more like scattering centers for the relevant low-wavelength phonons. The thermal conductivity at 300 K increases with $f$ despite the decrease owing to the addition of more percolated FLG conducting channels. The total cross-section of the percolated channels increases faster than the decrease in $c_p$.

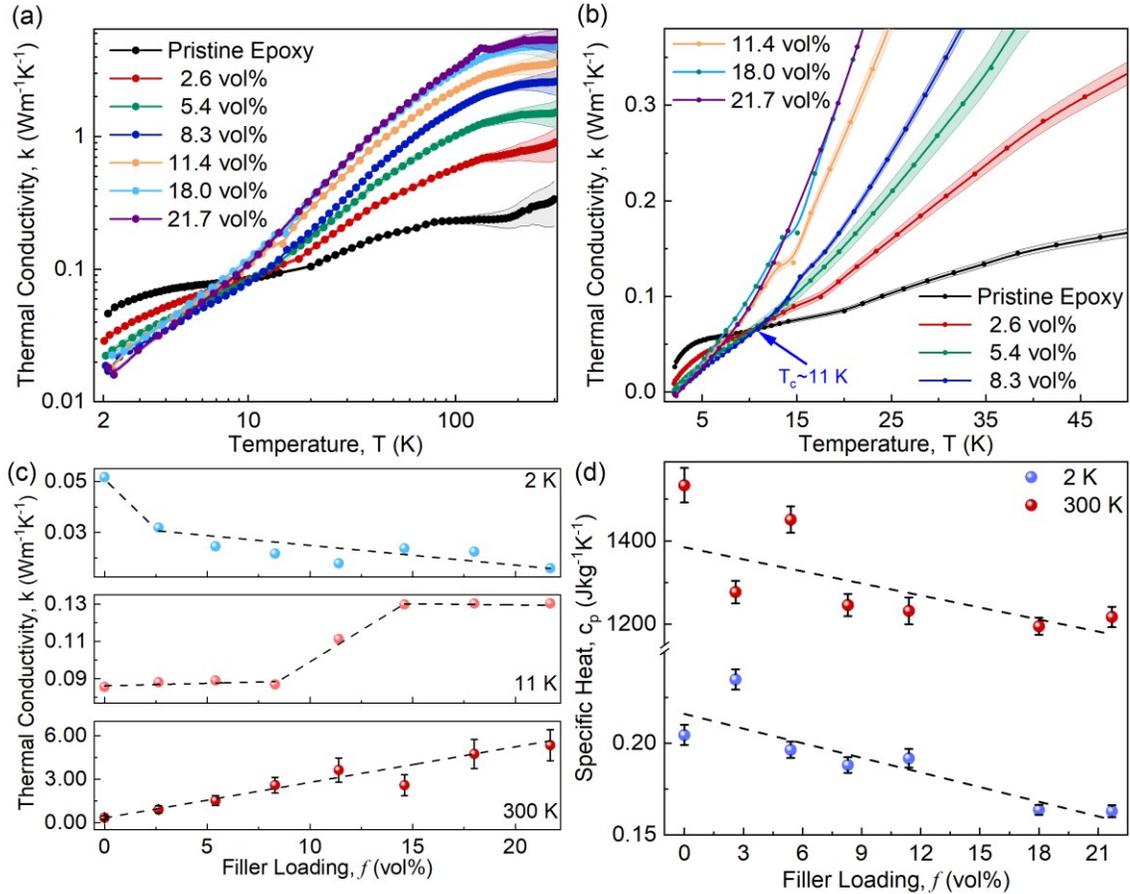

[**Figure 3: Temperature-dependent thermal transport characteristics of graphene composites**. (a) Thermal conductivity of graphene composites in the temperature range of 2 K ≤ $T$ ≤ 300 K in the log-log scale. The symbols are the experimental data points. The shaded area indicates the experimental uncertainty. (b) Thermal conductivity of the graphene composites in the low-temperature region, showing the cross-over temperature, $T_c$. (c) Thermal conductivity of the graphene composites as a function of the filler loading at 2 K, $T_c \sim 11$ K, and RT. (d) Specific heat of the graphene composites as a function of the filler loading at constant temperatures of 2 K and 300 K.]

**Effective medium model for cryogenic heat conduction for the low-loading composites:** We now develop a model of cryogenic heat conduction in graphene composites using Nan's effective medium model as the starting point.[55] According to this model, the effective thermal conductivity of a composite with randomly oriented low-loading fillers is given as[55]

$$k = k_m \frac{3 + f[2\beta_{11}(1 - L_{11}) + \beta_{33}(1 - L_{33})]}{3 - f[2\beta_{11}L_{11} + \beta_{33}L_{33}]}. \tag{1}$$

Here, $k_m$ is the thermal conductivity of the pristine epoxy, $k$ is the effective thermal conductivity of epoxy–graphene nanocomposite with the filler volume fraction $f$, $L_{ii}$ are the geometrical parameters that depend upon the aspect ratio, $p = t/L$, of graphene fillers with $t$ and $L$ being the thickness and lateral dimensions of the fillers. The details of $L_{ii}$ parameters and their definition can be found in the Supplementary Information. The parameters $\beta_{ii}$ contains information about the thermal boundary resistance at the filler-epoxy interface and is defined as:

$$\beta_{ii} = \frac{K_{ii}^c - k_m}{k_m + L_{ii}(K_{ii}^c - k_m)}, \tag{2}$$

where, $K_{ii}^c$ are the effective values of FLG thermal conductivity along different cartesian directions, that take into account the effect of interface thermal resistance. The effective thermal conductivities of the FLG fillers along the in-plane ($K_{11}^c \sim K_{22}^c$) and through-plane ($K_{33}^c$) are, respectively,

$$K_{11}^c = K_{22}^c = \frac{k_{in}}{1 + \gamma L_{11} k_{in}/k_m}, \tag{3}$$

$$K_{33}^c = \frac{k_{out}}{1 + \gamma L_{33} k_{out}/k_m}, \tag{4}$$

where, $\gamma = (1 + 2p)\alpha$ in which $\alpha = r_b k_m/t$ is a dimensionless parameter related to the interface thermal resistance, $r_b$, between the epoxy and filler, $k_{in}$ and $k_{out}$ represent the in-plane and through-plane thermal conductivity of pristine graphene fillers, respectively. To obtain the temperature dependence of the effective thermal conductivity, all involved parameters in Eq. (1) including $k_{in}$, $k_{out}$, $k_m$, and $r_b$ were taken to be temperature-dependent (see the Supplementary Information and Figure S3 for details). The combined effect of the high interfacial thermal resistance and low through-plane thermal conductivity of FLG at low temperatures (see Figure S4 (b) and Figure S3 (b)) results in a significantly small in-plane and through-plane "effective thermal conductivity" of FLG, $K_{11}^c$ and $K_{33}^c$, computed using Eq. (3) and (4). The results are shown in Figure 4 (a). The effective through-plane thermal conductivity of FLG becomes lower than that of the neat epoxy through the whole temperature range of 2 K up to 300 K. The graphene fillers oriented perpendicular to the heat flux serve as the extra thermal boundary resistance, a scattering center, rather than the conduit of heat. Note that $K_{33}^c$ is more than two orders of magnitude lower than the thermal conductivity of the pristine epoxy at 2 K. The in-plane thermal conductivity of FLG is only three times higher than that of the pristine epoxy at $T = 2$ K, whereas, at RT, its effective in-plane thermal conductivity is ~160 times higher than that of the neat epoxy.

The results of the calculations based on this effective medium model for composites with $f \leq$ 8.3 vol% are presented in Figure 4 (b) in the temperature range of 2 K up to 50 K. The model successfully reflects the experimental thermal conductivity characteristics for the low-concentration composites and correctly estimates the experimental cross-over temperature. Based on this model, we infer that the low effective through-plane thermal conductivity of FLG outweighs its highly effective in-plane thermal conductivity, causing the composite thermal conductivity to become lower than that of pure epoxy at low-temperature limits. As the loading of FLG increases, the effect of the low through-plane thermal conductivity of fillers dominates, resulting in decreasing the composite's thermal conductivity even more. In contrast, in high-temperature limits, both the in-plane and through-plane thermal conductivities increase, while simultaneously the interfacial TBR decreases as $r_b \sim T^{-3}$. This causes a rise in the effective thermal conductivity of the filler in both directions. At ~15 K, the effective thermal conductivities of FLG fillers recover enough, to result in a thermal transport enhancement. Above this temperature, the composite thermal conductivity becomes higher than that of the neat epoxy and it grows with increasing FLG loading. Thus, there is a transition in the thermal conductivity trend, with $k$ decreasing with increasing FLG concentration at low temperatures, and the opposite trend at higher temperatures. The latter leads to a cross-over temperature effect at 15 K for composite's $k$.

**Effective medium model for cryogenic heat conduction for the high-loading composites:** The considered effective medium model is unable to predict the thermal conductivity of the high-concentration composites since it does not include the effect of filler-filler contact. As the filler loading increases, the probability that fillers can physically contact each other within the host polymer grows. Increasing the filler content can result in entering into a strong thermal percolation regime.[21–23] In this regime, heat can conduct along a network of highly conductive connected FLG fillers within the polymer matrix. The disappearance of the cross-over temperature in high-concentration samples implies that the favorable effect of heat conduction along such networks might have overcome the negative effect of high interfacial thermal resistance at low temperatures. To predict the thermal conductivity characteristics at higher FLG concentrations, we use a recently introduced percolation-based effective medium model.[56] Through this model the composite thermal conductivity $k$ is determined by solving the following equation:

$$(1-f)\frac{k_0 - k}{k + \dfrac{k_0 - k}{3}} + \frac{f}{3}\left[\frac{2(k_{11} - k)}{k + S_{11}(k_0 - k)} + \frac{(k_{33} - k)}{k + S_{33}(k_{out} - k)}\right] = 0, \qquad (5)$$

where $k_{11}$ and $k_{33}$ are the effective in-plane and through-plane thermal conductivity of FLG fillers, which account for the interfacial thermal resistance. These parameters are calculated using the actual in-plane and through-plane thermal conductivities of FLG, $k_{in}$ and $k_{out}$. The shape parameters $S_{11}$ and $S_{33}$ are related to the aspect ratio of the FLG fillers, and $k_0$ is the thermal conductivity of an interlayer surrounding FLG fillers. This interlayer represents the interface thermal resistance surrounding the graphene fillers. Its role is to include the combined effect of interfacial thermal resistance at graphene-polymer and graphene-graphene contacts. The definitions of these parameters are explained in more detail in the Supplementary Information. The values of different parameters used in this model are listed in Table S1. Figure 4 (c) shows the results of the calculations for the high-concentration composites with $f \geq 11.4$ vol% along with the results of the low-loading effective medium model for the composites with $f \leq 8.3$ vol%. The predicted thermal conductivity in both cases is in good agreement with the experimental results (see Figure 3). One can conclude that the superior heat conduction along the percolated graphene fillers channel overcomes the negative effect of large interfacial thermal resistance at low temperatures causing the cross-over effect to disappear at compositions greater than $f \geq 11.4$ vol%.

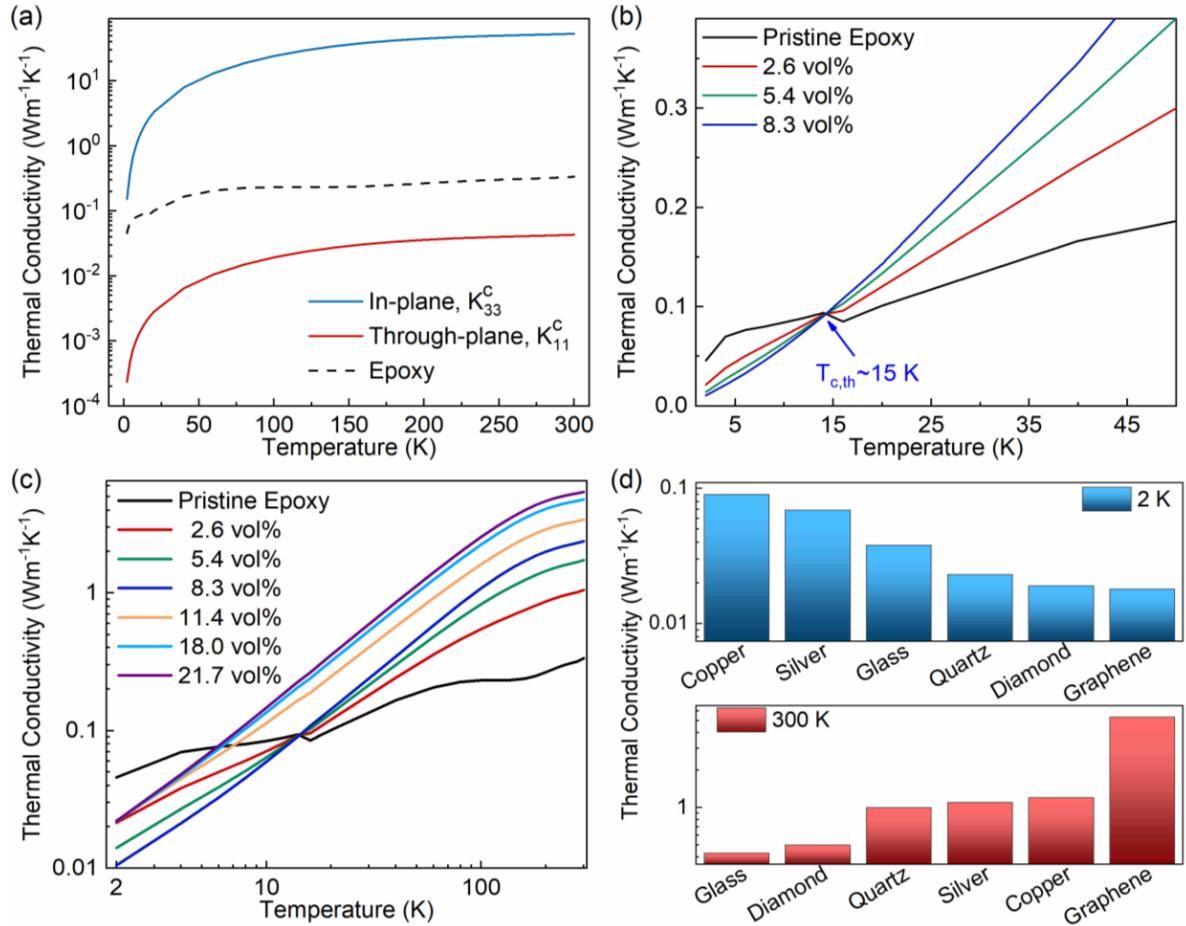

[**Figure 4: Thermal conductivity and thermal performance benchmarking.** (a) Calculated effective in-plane, $K_{33}^c$, and through-plane, $K_{11}^c$, thermal conductivity of graphene composites with FLG loading $f \leq 8.3$ vol%. (b) Calculated thermal conductivity of graphene composites with FLG loading of $f \leq 8.3$ vol%. The model successfully predicts the cross-over temperature, $T_{c,th} \sim 15$ K, which is in agreement with the experimental value. (c) Prediction of the percolative-based effective medium theory used to describe the thermal conductivity of composites with the high filler loading, $f \geq 11.4$ vol%. The data is presentred in the log-log scale together with the results of the low-loading model prediction presented in (b). (d) Comparison of the measured heat conduction properties of graphene composites with other materials. Note that graphene composites demonstrate better thermal insulation at cryogenic temperatures and superior thermal conductivity at RT. The experimental data for composites with other fillers are from Refs. [28,46].]

As follows from the above discussions, the thermal characteristics of semi-crystalline systems can drastically change in the low-temperature limits due to the increase in the dominant phonon wavelength and the changes in the phonon scattering mechanisms.[28,34,44] In this regime, fillers can act as phonon scattering centers rather than conductive inclusions, which suppresses the thermal transport of the composites even below the limit of its pristine amorphous polymer matrix.[28,34,44]

The temperature-dependent thermal transport data of composites shows a cross-over temperature, $T_c$, usually in the interval of 5 K to 20 K, at which the thermal conductivity of the composite is lower than the pristine polymer host at temperatures below $T_c$ and vice versa. The inclusion of fillers causes the plateau region observed in amorphous polymers to disappear. These two features are attributed to TBR at the interface of polymer-filler, which becomes dominant at low temperatures.[28] Cryogenic characteristics of graphene composites are unique in the sense that they offer the strongest suppression of the thermal conductivity below the cross-over temperature and the highest enhancement of the thermal conductivity above the cross-over temperature (see Figure 4 (d)). This is due to the atomic thickness of graphene and FLG, its geometry and exceptionally high intrinsic in-plane thermal conductivity. For practical applications, graphene composites offer dual functionality for the circuits and systems where both cooling and thermal insulation are required. At RT, the thermal conductivity of the graphene composite with $f = 21.7$ vol% reaches $6 \, \mathrm{Wm^{-1}K^{-1}}$, which is suitable for cooling semiconductor electronics, whereas a dilute composite with only 8.3 vol% filler loading reveals $k \sim 0.02 \, \mathrm{Wm^{-1}K^{-1}}$ at $T = 2$ K, providing excellent thermal insulating properties for superconducting electronics.

**CONCLUSIONS**

We demonstrated that at cryogenic temperatures, the thermal conductivity of graphene composites can be both higher and lower than that of the reference pristine epoxy, depending on the graphene filler loading and specific temperature. There exists a well-defined cross-over temperature – above it, the thermal conductivity increases with the addition of graphene; below it, the thermal conductivity decreases with the addition of graphene. The counter-intuitive trend was explained by the specificity of heat conduction at low temperatures. The randomly distributed graphene fillers can serve, simultaneously, as the scattering centers for acoustic phonons in the matrix material and as the conduits of heat. We also argued that the onset of the thermal percolation threshold can undergo modification owing to the dominance of the low-wavelength phonons that facilitate the filler-to-filler heat conduction even before the fillers are physically connected. The obtained results suggest the possibility of using composites with the same constituent materials for, both, removing the heat and thermally insulating electronic components at cryogenic temperatures. The latter is an important capability for the development of quantum computing technologies and cryogenically cooled conventional semiconductor electronics.

## CONTRIBUTIONS

F.K. and A.A.B. conceived the idea of the low-temperature study of graphene composites for thermal management applications in cryogenic electronics and quantum computing technologies. F.K. and A.A.B. coordinated the project and contributed to the experimental data analysis. Z.E.N. exfoliated graphene fillers, prepared the graphene composites, performed Raman measurements, and contributed to the thermal data analysis. Y.X. conducted specific heat and thermal conductivity measurements using PPMS. J.B. assisted with the thermal data analysis. J.G. performed numerical modeling of thermal conductivity and contributed to data analysis. X.C. supervised thermal measurements and contributed to the experimental data analysis. F.K. and A.A.B. led the manuscript preparation. All authors contributed to the writing and editing of the manuscript.

## Acknowledgements

A.A.B. acknowledges the support of the Vannevar Bush Faculty Fellowship from the Office of Secretary of Defense (OSD), under the office of Naval Research (ONR) contract N00014-21-1-2947.

**Supporting information**

# Low-Temperature Thermal Characteristics of Graphene Composites – Applications in Thermal Management for Quantum Technologies

Zahra Ebrahim Nataj,[1] Youming Xu,[2] Jonas Brown,[1] Jivtesh Garg,[3] Xi Chen, [2] Fariborz Kargar,[1,‡] and Alexander A. Balandin[1,*]

[1]Nano-Device Laboratory and Phonon Optimized Engineered Materials Center, Department of Electrical and Computer Engineering, University of California, Riverside, California 92521 USA

[2]Department of Electrical and Computer Engineering, University of California, Riverside, California 92521 USA

[3] Department of Aerospace and Mechanical Engineering, University of Oklahoma, Norman, Oklahoma 73019 USA

---

‡ Corresponding authors: fkargar@ece.ucr.edu ; balandin@ece.ucr.edu ; web-site: http://balandingroup.ucr.edu/

## Contents

# 1    Sample Preparation

For each composite sample with desired Few-layer graphene (FLG) loading, commonly produced FLG (xGnP H-25, XG Sciences, US) with an average surface area of $65 \ m^2 g^{-1}$ and average lateral dimensions of 25 $\mu m$ were mixed with the base epoxy resin (bisphenol A (epichlorhydrin); molecular weight, 700; Allied HighTech Products, Inc.) at predetermined proportions. In order to have a uniform compound, FLG was added in several steps and mixed for 3 minutes at 800 rpm in a high-shear speed mixer in each step. The hardener (triethylenetetramine, Allied HighTech Products, Inc.) was then added at a mass ratio of 12:100 to the epoxy resin. The final compound was mixed and vacuumed for 10 minutes to remove trapped air bubbles. This procedure was performed three times to achieve a void-free combination. Then, the samples set aside at room temperature for about 8 hours to harden. At higher graphene concentrations (from 20 wt% and above), it was hard to reduce its porosity with just mixing and vacuuming the compound. So, one step was added to the previous procedures. Before leaving the samples at room temperature to be rigid, the completed mixture was poured into spherical molds, gently squeezed. Finally, all the samples were put in a 130℃ furnace for 3 hours to cure and solidify. The final composite samples were disks with a diameter of 25.4 mm and a thickness of 5 mm. the optical images of samples are shown in Figure 1.

## 2    Mass Density Measurement

The mass density of the samples was determined using the Archimedes principle and an electronic scale (Mettler Toledo). The density is calculated using the formula,

$$\rho_c = (w_a/(w_a - w_w)) \times (\rho_w - \rho_a) + \rho_a \tag{S1}$$

where $w_a$, $w_w$ are the sample's weight in air and water, respectively, and $\rho_w$ and $\rho_a$ are the density of ionized water and air ($0.0012\ gcm^{-3}$) at room temperature .The results of the mass density measurements for different filler concentrations presented in Figure S1 indicate that addition of graphene fillers to the composites would increase the density of the samples.

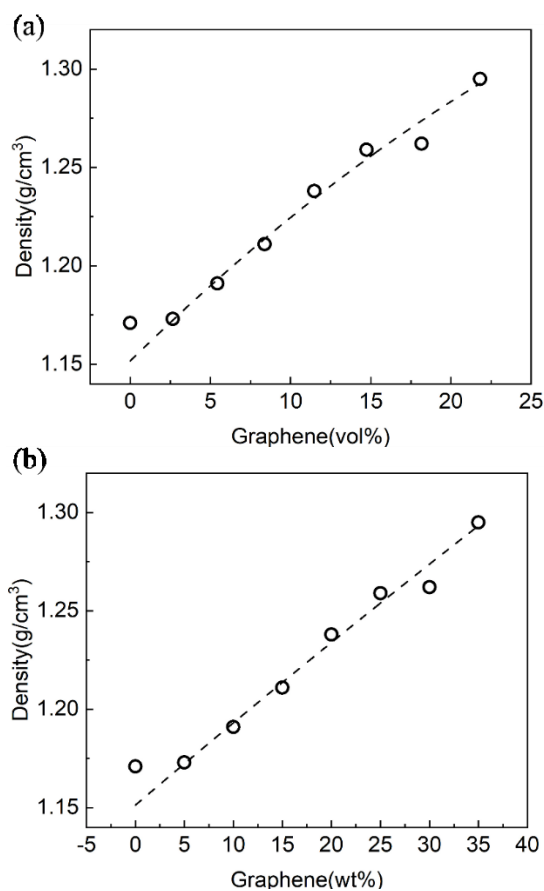

Figure S1: Mass density of composites as a function of, a) volume loading and b) mass loading fraction of graphene.

# 3    Thermal Conductivity Measurement and uncertainty analysis

The samples were cut into a typical dimension of 1×1×10 mm for the thermal conductivity measurement. A Quantum Design Physical Property Measurement System (PPMS) was employed to measure the thermal conductivity from 2 to 300 K with a steady-state 4-probe continuous mode. In this mode, the temperature sweeps between 2 and 300 K at low rate (0.3 K/min) and measurements are taken continually, while the measurement parameters like heating power and period are adjusted by software automatically.[1] When a square-wave heat pulse is applied, the temperature difference between the hot probe and cold probe is measured as a function of time and a steady-state temperature difference is fitted using the following equation:

$$\Delta T = \Delta T_\infty \times (1 - \frac{\tau_1 \times \exp\left(-\frac{t}{\tau_1}\right) - \tau_2 \times \exp\left(-\frac{t}{\tau_2}\right)}{\tau_1 - \tau_2}) \tag{S2}$$

where $\Delta T$ is the measured temperature difference, $\Delta T_\infty$ is the steady-state temperature difference, $\tau_1$ and $\tau_2$ are time constants and can be obtained from the fitting.

When $\Delta T_\infty$ is calculated, the thermal conductance $K$ is calculated as

$$K = P/\Delta T_\infty \tag{S3}$$

where $P$ is the heat flowing through the sample. And $P$ is calculated using Joule heat minus radiation heat:

$$P = I^2 R - P_{rad} \tag{S4}$$

where $I$ is the current flowing into the sample and $R$ is the resistance of the heater, the radiation loss is calculated as

$$P_{rad} = \sigma_T \times (\frac{S}{2}) \times \varepsilon \times (T_{hot}^4 - T_{cold}^4) \tag{S5}$$

where $\sigma_T = 5.67 \times 10^{-8}$ Wm$^{-2}$K$^{-4}$ is the Stefan-Boltzmann constant, $S$ is the surface area of the sample, $\varepsilon$ is the emissivity, $T_{hot}$ and $T_{cold}$ are the temperature of hot probe and cold probe.

Then the shoe assembly's thermal conductance $K_{shoe}$ is deduced from $K$ to get sample thermal conductance $K_{sample}$ as:

$$K_{sample} = K - K_{shoe} \tag{S6}$$

$$K_{shoe} = aT + bT^2 + cT^3 \tag{S7}$$

where $a$, $b$, $c$ are constants. The thermal conductivity of the sample is calculated as:

$$\kappa = K_{sample} \times l/(w \times t) \tag{S8}$$

where $l$, $w$, $t$ are the distance between the hot and cold probe, width and thickness of the sample, respectively.

The error of the thermal conductivity measurement includes the fitting of $\Delta T$, the error of the heating power, the error in the estimation of the sample radiation due to the error of sample surface area and emissivity, the error in $K_{shoe}$, the error from the measurements of the size of sample and the error from the measurement of the distance between hot and cold probes $d$. So the error is calculated as

$$\sigma(\kappa) = \kappa \times \sqrt{\left(\frac{R_{\Delta T}}{\Delta T_\infty}\right)^2 + \left(2\frac{IR\partial I}{P}\right)^2 + \left(\frac{0.2 \times P_{rad}}{P}\right)^2 + \left(\frac{0.1 \times T_\infty \times K_{shoe}}{P}\right)^2 + \left(\frac{\Delta l}{l}\right)^2 + \left(\frac{\Delta w}{w}\right)^2 + \left(\frac{\Delta t}{t}\right)^2} \tag{S9}$$

where $R_{\Delta T}$ is the residual term from the $\Delta T$ versus $t$ fitting.

## 4   Heat capacity measurement and uncertainty analysis

The sample of a typical mass of 5 mg was used for the heat capacity measurement from 2 to 300 K with the PPMS. The measurement of heat capacity assumes that the temperature of the sample $T_s$ is different from the temperature of the platform $T_p$ because the thermal contact between the sample and the platform is not good enough.[2] The heat transfer between the sample and the platform is described by the following equations:

$$C_{platform}\frac{dT_p}{dx} = P(t) - K_w\big(T_p(t) - T_b\big) + K_g(T_s(t) - T_p(t)) \tag{S10}$$

$$C_{sample}\frac{dT_s}{dx} = -K_g(T_s(t) - T_p(t)) \tag{S11}$$

where $P(t)$ is the heater power. $C_{platform}$ and $C_{sample}$ are the heat capacity of the platform and the sample, respectively. $K_w$ is the thermal conductance of the supporting wires, $K_g$ is the thermal conductance between the platform and the sample due to the grease. $T_b$ is the temperature of the thermal bath. $T_s$ and $T_p$ are function of time.

The error of the heat capacity measurements includes the total heat capacity error from the fitting of the heat capacity parameters, the addenda heat capacity error from the addenda measurement, sample mass error and the fitting deviation error from the modeling. Therefore, the error of heat capacity $C_p$ is calculated as

$$\sigma\big(C_p\big) = C_p \times \sqrt{(\frac{R_{C_{total}}}{C_{total}})^2 + (\frac{R_{C_{platform}}}{C_{platform}})^2 + (\frac{\Delta M}{M})^2 + (\frac{R_{\Delta T}}{\Delta T})^2} \tag{S12}$$

where $R_{C_{total}}$ and $R_{C_{platform}}$ are the mean-square deviation of the fit to variations in the total heat capacity and platform heat capacity, $M$ is the mass of sample and $R_{\Delta T}$ is the fitting deviation error.

Figure S2 shows the heat capacity of all samples as a function of temperature in the range of $100 \text{ K} \leq T \leq 300 \text{ K}$. Based on the figure, one can see that heat capacity of each filler concentration increases as temperatures increase with a linear relation. Also, by increasing the filler amount the heat capacity decreased. However, there is an irregularity in this behavior for filler concentration of 5.4 vol% which might be because of systematic errors.

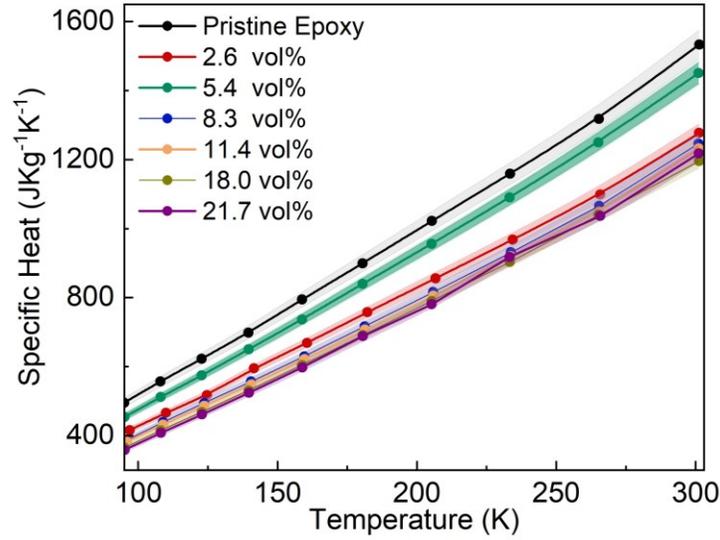

Figure S2: Heat capacity of composites with different graphene loading as a function of temperature. The shaded area around the experimental data points displays the errors involved in the measurements.

## 5    Effective medium model for cryogenic heat conduction for low-loading composites

In the equation (1) of the main text, $L_{ii}$ are geometrical parameters and depend upon the aspect ratio, $p = t/L$, of graphene fillers with $t$ and $L$ being the thickness and lateral dimensions of fillers. For oblate inclusions such as nanoplatelets, where $p < 1$, these geometrical parameters, $L_{ii}$, are computed using the following equations,

$$L_{11} = L_{22} = \frac{p^2}{2(p^2-1)} + \frac{p}{2(1-p^2)^{3/2}} \cos^{-1} p \qquad (S13)$$

$$L_{33} = 1 - 2L_{11} \qquad (S14)$$

**Temperature Dependence of thermal conductivity of graphite:**

Temperature dependent in-plane thermal conductivity of natural graphite was obtained from [3]. Temperature dependent through-plane thermal conductivity was obtained from [4]. These data are presented in Figures S3a and S3b. It is visible that thermal conductivity of graphite can reach very low values at low temperatures. At 4 K, the in-plane thermal conductivity of graphite decreases to ~ 1 W/mK, while the through-plane thermal conductivity reaches 0.23 W/mK.

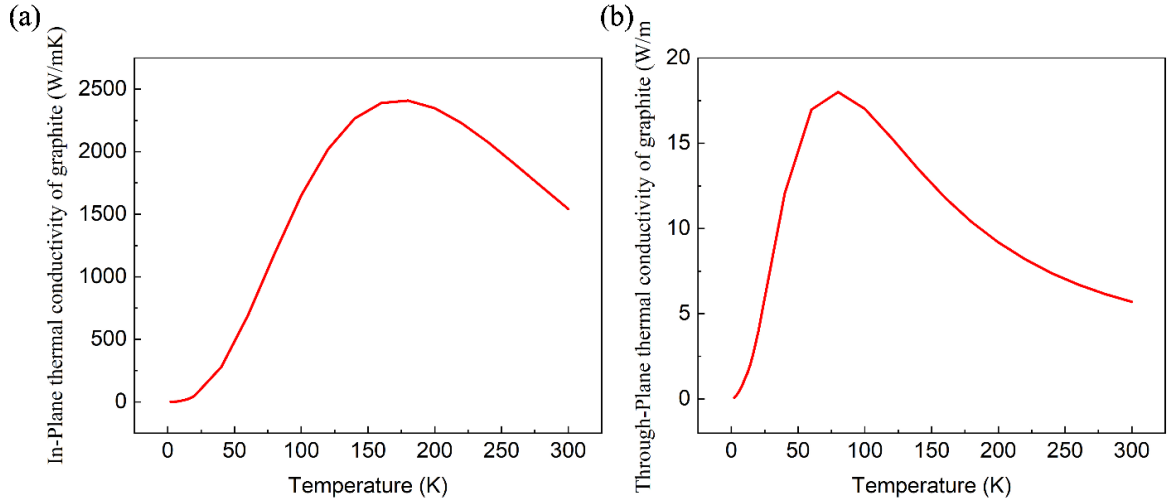

Figure S3: Temperature dependence of, a) in-plane and b) through-plane thermal conductivities of natural graphite.

The explicit temperature dependence of in-plane and through-plane thermal conductivities (fitted to the data in Figs. S3a and b) is given by following equations:

$$k_{in} = \exp(0.00116[\ln(T)]^6 - 0.02386[\ln(T)]^5 + 0.17360[\ln(T)]^4 - 0.66809[\ln(T)]^3 + 1.67342[\ln(T)]^2 - 0.19473[\ln(T)]^1 - 1.65861)$$

(S15)

$$k_{out} = \exp(0.0975538[\ln(T)]^4 - 1.6222796[\ln(T)]^3 + 9.1860571[\ln(T)]^2 - 19.9450584[\ln(T)]^1 + 14.4337754)$$

(S16)

The thermal conductivity of epoxy was taken to be temperature dependent from measurements. Following curve fits can provide good fit to the experimental data.

For  2 K< T <14 K

$$k_{epoxy} = 1.5044 \times 10^{-6}T^5 - 7.2195 \times 10^{-5}T^4 + 1.3521 \times 10^{-3}T^3 - 1.2278 \times 10^{-2}T^2 + 5.5773 \times 10^{-2}T - 2.6347 \times 10^{-2}$$

(S17)

For 14 K < T < 300 K

$$k_{epoxy} = 1.3105 \times 10^{-14}T^6 - 1.1288 \times 10^{-11}T^5 + 3.3133 \times 10^{-9}T^4 - 3.08119 \times 10^{-7}T^3 - 2.01387 \times 10^{-5}T^2 + 4.9861 \times 10^{-3}T + 1.13744 \times 10^{-2}$$

(S18)

Explicit temperature dependence of interface thermal resistance is provided below.

$$R_{interface} = \exp(0.0111[\ln(T)]^4 - 0.1147[\ln(T)]^3 + 0.3702[\ln(T)]^2 - 1.5114[\ln(T)]^1 - 8.7497)$$

(S19)

The interface thermal conductance and thermal resistance (inverse of conductance) is shown in Figure S4.

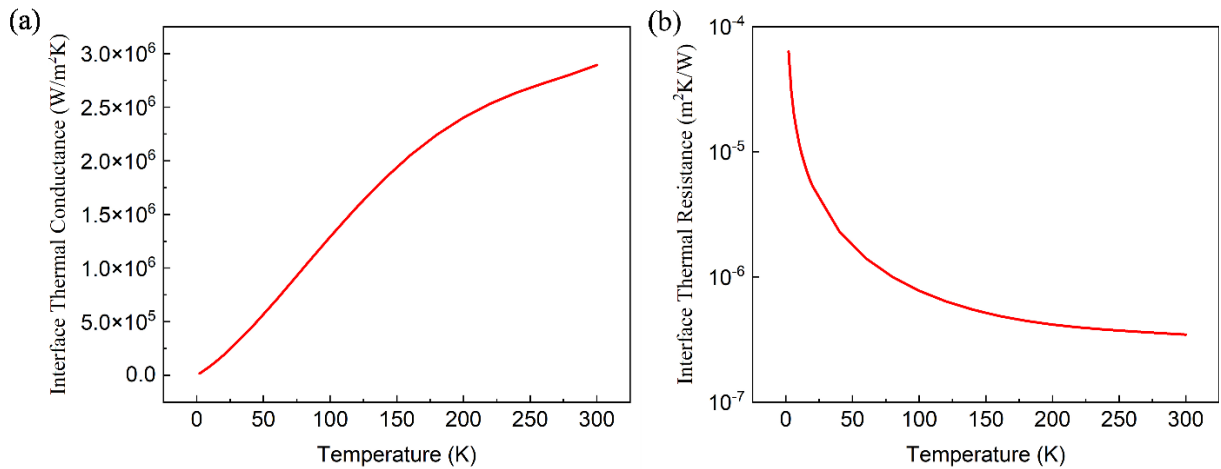

Figure S4: Temperature dependence of interfacial, a) thermal conductance and b) thermal resistance, between graphene and pure epoxy.

## 6   Effective medium model for cryogenic heat conduction for high-loading composites (including percolation effect)

In the equation (5) of main text, $S_{11}$ and $S_{33}$ are the shape parameters related to the aspect ratio of graphitic nanosheets, given by following equations.

$$S_{11} = S_{22} = \frac{p}{2(1-p^2)^{3/2}}\left[\cos^{-1} p - p(1-p^2)^{1/2}\right], \qquad p < 1 \tag{S20}$$

$$S_{33} = 1 - S_{11} \tag{S21}$$

In the above equations, $p$ is the aspect ratio of the fillers. The effective in-plane and through-plane thermal conductivities, $k_{11}$ and $k_{33}$ (in Eq. 5 in main text) are computed using

$$k_{11} = k_0\left[1 + \frac{(1-c_{int})(k_{in} - k_0)}{c_{int}S_{11}(k_{in} - k_0) + k_0}\right] \tag{S22}$$

$$k_{33} = k_0\left[1 + \frac{(1-c_{int})(k_{out} - k_0)}{c_{int}S_{11}(k_{out} - k_0) + k_0}\right] \tag{S23}$$

In above equations, $k_0$ is thermal conductivity of an interlayer surrounding graphene sheets. This interlayer represents the interface thermal resistance surrounding the graphene particles. The interlayer is used to represent the combined effect of graphene-epoxy and graphene-graphene contact resistance as described in the article. The values of different parameters used in this percolation based effective medium model are described below. Once again, wherever appropriate, we used temperature dependent properties. Table S1 below shows the values of different parameters used in the above calculations.

Table S1: Thermal conductivity calculations parameters

| Material Parameters | Values |
|---|---|
| Average graphene lateral length, $l$, | 15 μm |
| Average graphene thickness | 15 nm |
| Aspect ratio of the graphene filler | 0.001 |
| Thermal conductivity of epoxy | Temperature dependent |
| Thermal conductivity of graphene filler, $k_1$ and $k_3$ (W/mK) | Temperature dependent |
| Thermal conductivity of interlayer with Kapitza resistance | Temperature dependent |
| Thermal conductivity of the interlayer with a firmly developed graphene-graphene contact state, | Temperature dependent |

The in-plane ($k_{in}$) and through-plane thermal conductivities ($k_{out}$) of graphene were taken to betemperature dependent as shown in Figs. S3a and b. . The combined interfacial resistance of graphene/epoxy and graphene/graphene contactis modeled as an interlayer in above theory with an effective thermal conductivity of $k_0$ which is taken to be weighted sum of thermal conductivity of graphene/epoxy and graphene/graphene contact. The thickness of this interlayer was nominally taken to be 1.0 nm. Thermal conductivity of the interlayer based on graphene-epoxy contact was computed from the thickness and interfacial resistance of graphene/epoxy contact which is taken to the same as given by equation (S7). The thermal conductivity of interlayer based on graphene/graphene contact was computed from the interfacial resistance at graphene-graphene contact which was was taken to be lower than graphene-epoxy interfacial thermal resistance by a factor ranging from 4.5 for 11.4 vol% composition to 6.0 for 21.7 vol% composition.